\documentclass[journal]{IEEEtran}  

\usepackage{graphicx}
\usepackage[caption=false]{subfig}
\usepackage{multirow}
\usepackage{amsmath}

\graphicspath{{./Figures/}}

\DeclareGraphicsExtensions{.eps}

\title{
	A Low-Power Accelerator for Deep Neural Networks with Enlarged Near-Zero Sparsity 
}

\author{Yuxiang~Huan,~\IEEEmembership{Student Member,~IEEE,}
	Yifan~Qin,~
	Yantian~You,~
	Lirong~Zheng,~\IEEEmembership{Senior Member,~IEEE}
	and~Zhuo~Zou,~\IEEEmembership{Member,~IEEE,}

\thanks{Y. Huan, Z. Zou, and L. Zheng is with the Fudan University, Shanghai, China, and also with the Royal Institute of Technology (KTH), SE 164-40 Kista, Sweden (Contacts: Z. Zou: zhuo@kth.se, L. Zheng: lrzheng@fudan.edu.cn).}
\thanks{Y. Qin is with the Fudan University, Shanghai, China.}

\thanks{Y. You is with the Royal Institute of Technology (KTH), SE 164-40 Kista, Sweden.}

}

\begin{document}
	
\abovedisplayskip=5.0pt plus 2.0pt minus 2.0pt
\belowdisplayskip=5.0pt plus 2.0pt minus 2.0pt	
	\maketitle


\begin{abstract}

	It remains a challenge to run Deep Learning in devices with stringent power budget in the Internet-of-Things. This paper presents a low-power accelerator for processing Deep Neural Networks in the embedded devices. The power reduction is realized by avoiding multiplications of near-zero valued data. The near-zero approximation and a dedicated Near-Zero Approximation Unit (NZAU) are proposed to predict and skip the near-zero multiplications under certain thresholds. Compared with skipping zero-valued computations, our design achieves 1.92X and 1.51X further reduction of the total multiplications in LeNet-5 and Alexnet respectively, with negligible lose of accuracy. In the proposed accelerator, 256 multipliers are grouped into 16 independent Processing Lanes (PL) to support up to 16 neuron activations simultaneously. With the help of data pre-processing and buffering in each PL, multipliers can be clock-gated in most of the time even the data is excessively streaming in. Designed and simulated in UMC 65 nm process, the accelerator operating at 500 MHz is $>$ 4X faster than the mobile GPU Tegra K1 in processing the fully-connected layer FC8 of Alexnet, while consuming 717X less energy.
		
\end{abstract}
	
	

	\section{INTRODUCTION}
	
\IEEEPARstart{I}{t} is an inevitable trend that trillions of devices with sensing and processing capabilities will be connected to form the Internet-of-Things (IoT). Beyond today's IoT systems that are based on Radio Frequency IDentifications (RFIDs) and Wireless sensor networks (WSNs), recent advances in low power design and artificial intelligence help to shift the functionalities of devices from sensing to perception. This paradigm shift requires ubiquitously embedded nodes to be capable of extracting features or patterns from sensed raw data, so as to enable intelligent perception of the surrounding world. The emergence of Deep Neural Network (DNN) models make it feasible, as they exhibit superior performance over traditional approaches in learning of data. Due to the complexity of the DNNs, they require high performance platforms, like CPU, GPU \cite{ISCAS14, icml2013}, or FPGA \cite{ISCA10,CVPR14}, and thus are usually run on high performance servers or clusters. Rather than sending raw data to the cloud, local processing on-device is more preferred in terms of security, communication overhead, and delay. It is beneficial that DNN models can be optimized to fit in resource-constrained IoT devices \cite{ASPDAC16}.
	
To meet the requirements of IoT devices, complex DNN models should be processed in a more compact manner as most of them are heavily over-parameterized. ReLU activation \cite{alexnet} and network pruning \cite{hanNIPS15} are thus widely used to simplify the inference of the DNN, enlarging the sparsity of the neural input data and the weights in most of DNN layers. Computation of sparse data can be omitted, as it contributes negligible efforts to the final results but occupies the limited processing resources. Existing platforms, such as CPU, DSP, or GPU, are inefficient to run the sparse model due to the weak support for sparse data detection and skipping. In addition, they suffer from high energy consumption and cost, which doesn't meet power and cost constraints of IoT devices. Specialized designs have been implemented to avoid the multiplications of the sparse data  \cite{eyeriss,EIE,VLSI16}, yet the sparsity is only restricted to zero-valued data. 


This paper extends the zero-valued sparsity to near-zero-valued sparsity, in order to avoid complex multiplications of near-zero valued data through approximation. A corresponding Near-zero Approximation Unit (NZAU) is proposed to predict and skip near-zero multiplications under certain thresholds. Based on the NZAU, a low-power accelerator with 256 multipliers is designed for energy-efficient processing of DNNs. By adjusting the threshold of the NZAU, the design achieves a 1.92X and 1.48X further reduction of the total multiplications than zero-skipping in LeNet-5 \cite{lenet} and Alexnet \cite{alexnet}, respectively. With the help of data pre-processing and buffering in each Processing Lane, the multipliers can be clock-gated in most of the time with no stall in processing. Implemented in UMC 65 nm process, the accelerator operating at 500MHz is $>$ 4X faster than the mobile GPU Tegra K1 in processing the fully-connected layer FC8 of Alexnet \cite{alexnet}, while consuming 717X less energy.

The remainder of this paper is organized as follows. Section II gives the mechanism of near-zero approximation. Section III describes the implementation details of the proposed accelerator. The near-zero approximation and the performance and energy consumption of the accelerator are evaluated and discussed in Section IV. Finally, conclusions are drawn in Section V.
  		
\section{Approximation Mechanism}

\begin{figure*}[!t]
\centering
\includegraphics[width=0.80\textwidth]{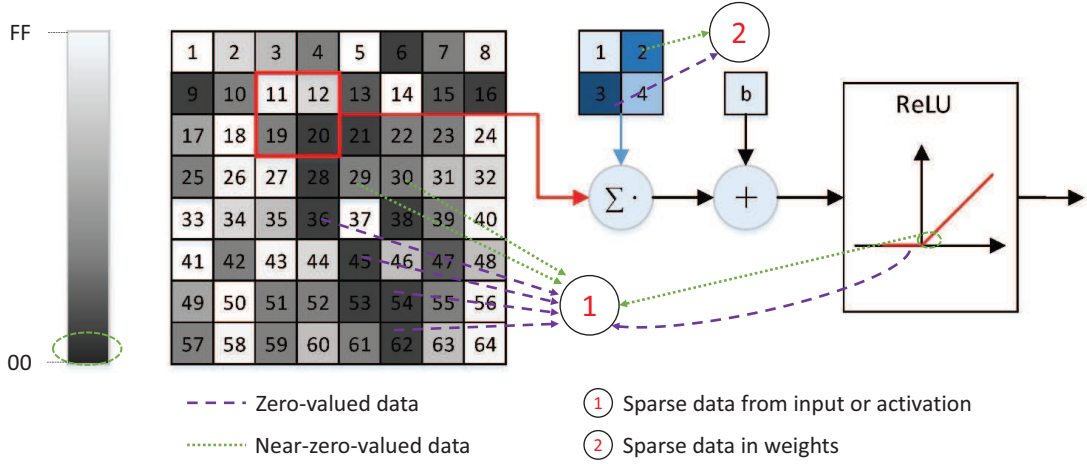}
\caption{Zero-valued sparsity and near-zero-valued sparsity in the inference of DNN model}
\label{fig_sparse}
\end{figure*}

As a key part of the Deep Neural Network, matrix-vector multiplications are commonly used to compute the activation of a neuron. The activation usually follows the (\ref{eq:1}) (biases are omitted here). 
\begin{small}
	\begin{equation}
	\label{eq:1}
	y_{i} = ReLU\left ( \sum_{j=0}^{N}\left ( W_{i,j} * x_{j} \right )\right )
	\end{equation}
\end{small}	
Existence of sparsity in both the weight matrix $W$ and the input vector $x$ implies opportunities of optimizing the complex matrix-vector multiplication. As depicted in Fig. \ref{fig_sparse}, the sparsity of the input vector may result from the original input data or the activation by ReLU, while the weight matrix becomes sparse if the network pruning is used. 

\subsection{Near-Zero-valued Sparsity}	

%


%
%
%
%
Though skipping multiplications of the sparse data brings benefits for improving energy efficiency, the sparsity is only restricted to zero-valued data. Due to the resilience nature of the neural network, small error is tolerable during the inference of the network. It is possible to discard the multiplications whose results are so small that cause little effect on the final results. As indicated by (\ref{eq:2}), (\ref{eq:3}), (\ref{eq:4}) and (\ref{eq:5}), the computation of (\ref{eq:1}) can therefore be divided into three parts: 

\begin{small}
	\begin{align}
	y_{i} &= ReLU\left ( P1 + P2 + P3\right )\approx ReLU\left ( P1\right )  \label{eq:2}\\	
	P1 &= \sum_{j\in \left \{W_{i,j} * x_{j}\neq 0 \right \}}^{ }\left ( W_{i,j} * x_{j} \right )  \label{eq:3}\\
	P2 &= \sum_{j\in \left \{W_{i,j} * x_{j}= 0 \right\}}^{ }\left ( W_{i,j} * x_{j} \right )  \label{eq:4}\\
	P3 &= \sum_{j\in \left \{W_{i,j} * x_{j}\approx 0 \right\}}^{ }\left ( W_{i,j} * x_{j}  \label{eq:5}\right ) 
	\end{align}
\end{small}

P1 refers to the part of multiplications that result in none-zero values; P2 refers to the part of multiplications that result in zero value and can be bypassed through zero-skipping techniques; P3 refers to the part of multiplications that can be approximated to zero if constrained by a certain threshold. As both of the P2 and the P3 are discarded if a proper threshold is used, data involved in the P3 can be regarded as "sparse". To distinguish the sparsity that only covers the zero-valued data, we extend the sparsity to include the portion of multiplications in P3 and define new sparsity as near-zero-valued sparsity (NZ-sparsity). Therefore, large scale matrix-vector multiplications can be further optimized if near-zero multiplications are effectively handled. 
	
\subsection{Near-Zero Approximation}
For a given threshold of a multiplication, value ranges of the two operands are dynamic. There are no restrictions directly applied to the operands. Therefore, constraining the result of the multiplication is more beneficial than constraining the operands, as the first one allows wider value ranges of the operands. In this case, the result of a multiplication should be predicted without doing the complex floating-point or fixed-point computation. Hence, we devise an approach to predict and approximate the result of a multiplication based on detecting the most significant bits of the operands.


%

	\begin{figure*}[!t]
		\centering
		\includegraphics[width=0.95\textwidth]{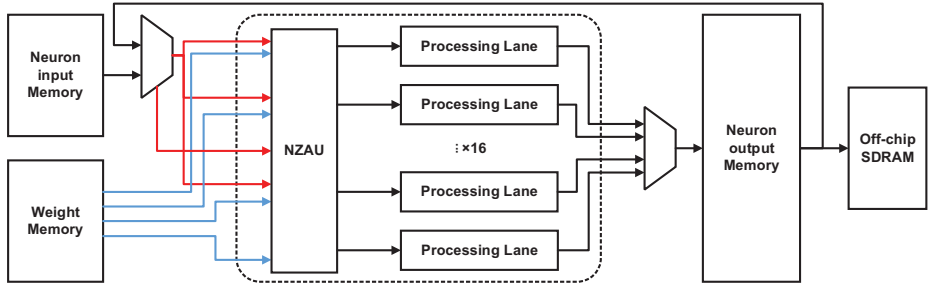}
		\caption{Architecture of the proposed accelerator: the Near-Zero Approximation Unit pre-processes the excessive streaming data, while the 16 Processing Lanes are responsible for buffering and computation of the operands that can not be skipped.}
		\label{fig_arch}
	\end{figure*}
	
	\begin{figure*}[!t]
		\centering
			\subfloat[The Near-Zero Approximation Unit that pre-processes 16 multiplications simultaneously]{%
			\includegraphics[width=0.35\textwidth]{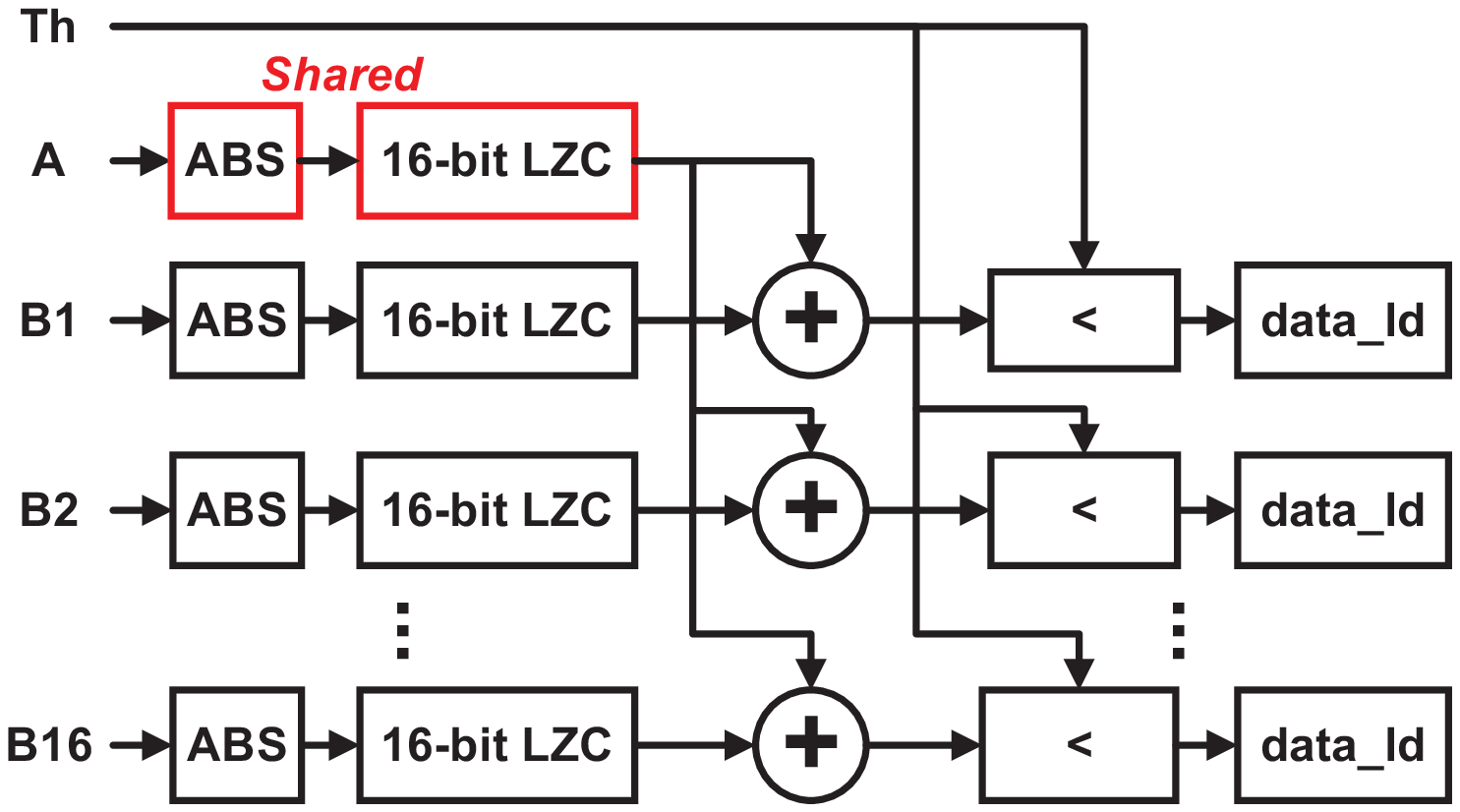}%
			\label{fig:nzau}%
		}\qquad
		\subfloat[Architecture of the Processing Lane: both of the buffering circuits and the computational units are clock-gated, and the operands registers isolate the computational units from the local buffers, halting the processing when assembling the operands.]{%
			\includegraphics[width=0.6\textwidth]{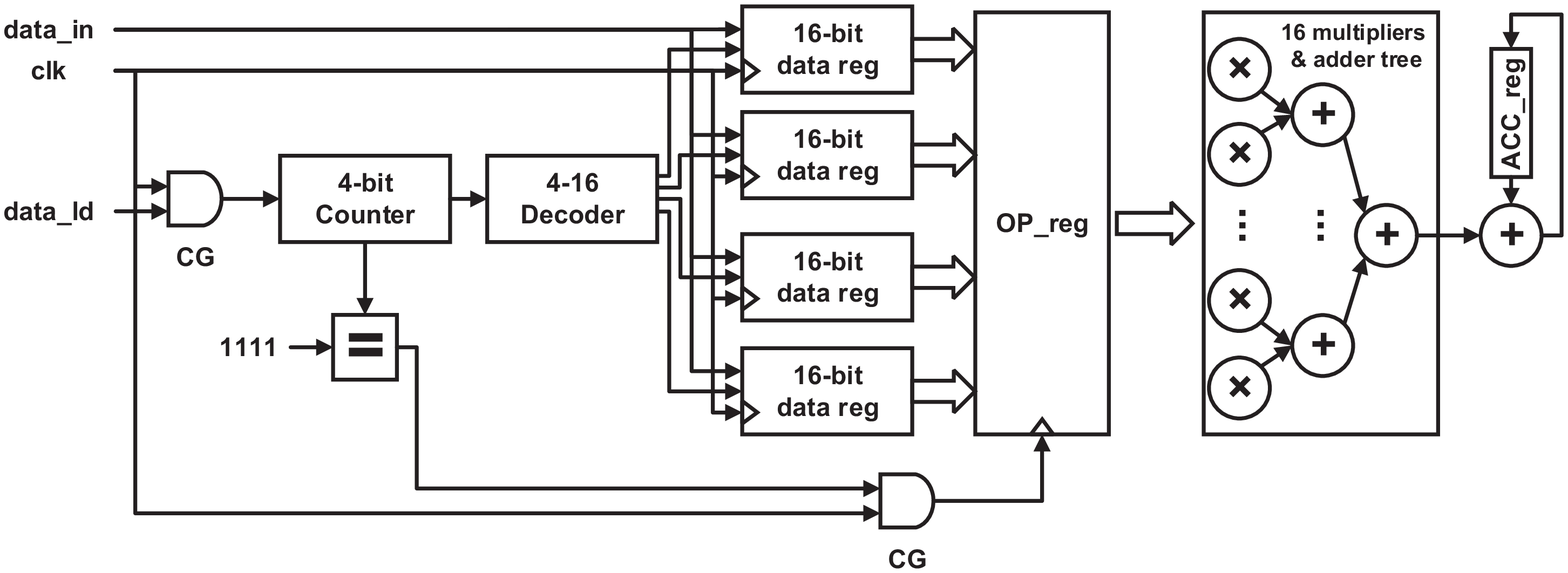}%
			\label{fig:pl}%
		}
		\caption{Key building blocks of the proposed accelerator}
	\end{figure*}


    
The most significant bit of the data is a good indicator for its relative value, which can be determined through leading zero detection. Assuming that there are two N-bit operands, which have the leading zero counts as $l_{A}$ and $l_{B}$ correspondingly. Binary numbers that have the same leading zero count as $l_{A}$ or $l_{B}$ lie in the ranges indicted as: 

\begin{small}
    \begin{equation}
    \label{eq:6}
	2^{N-l_{A}-1}\leq A< 2^{N-l_{A}}, 2^{N-l_{B}-1}\leq B< 2^{N-l_{B}}
    \end{equation}
\end{small}	
When multiplying the two operands, the result is then expressed as: 
\begin{small}
    \begin{equation}
    \label{eq:7}
    2^{2N-\left (  l_{A}+l_{B}\right )-2}\leq A\cdot B< 2^{2N-\left (  l_{A}+l_{B}\right )}
    \end{equation}
\end{small}	  
If applying the function of leading zero counting (denoted as $f_{l}$)) to (\ref{eq:7}), it can be re-written as:
\begin{small}
	\begin{align}
	f_{l}(2^{2N-\left (  l_{A}+l_{B}\right )}) <f_{l}(A\cdot B) \leq f_{l}(2^{2N-\left (  l_{A}+l_{B}\right )-2})   \label{eq:8}
	\end{align}
\end{small}	
it can be further inferred as:
\begin{small}
	\begin{align}
		\Rightarrow  l_{A}+l_{B}-1 <f_{l}(A\cdot B) \leq l_{A}+l_{B} +1 \label{eq:9}
	\end{align}
\end{small}	
It can be found that the leading zeros of the result is either $l_{A} + l_{B}$ or $l_{A} + l_{B} + 1$. In other words, the total leading zeros of the two operands can determine the relative value of the result. Based on this, constraining the results can be turned into constraining the total leading zeros of the operands (denoted as $l_{total}$). If $l_{total}$ is larger than the given threshold, the operands will be discarded and the multiplication result will be approximated to 0. Otherwise, these operands will be kept for further multiplication. 

The effectiveness of such near-zero approximation has been preliminarily examined in a handcrafted Neural Network in our previous work \cite{HYX_SOCC16}. In this work, we apply the approximation mechanism to two more complex models: LeNet-5 and Alexnet. Comparing with skipping zero-valued multiplications, this mechanism further reduces of the total multiplications of the two models by 1.92X and 1.51X respectively with negligible lose of accuracy. Detailed discussion and evaluation will be given in Section III.

	
\section{Hardware implementation}
As most of the existing platforms, such as CPU or GPU, are inefficient to perform the proposed near-zero approximation, an low-power accelerator for matrix-vector multiplications is implemented with dedicated support of the proposed approximation mechanism. 
	
The architecture of the accelerator is illustrated in Fig. \ref{fig_arch}. This accelerator adopts a SIMD-like architecture to enable efficient parallel operations of the data. The computational units of the accelerator are consisting of 256 multipliers and 16 adder trees. To minimize the overhead of memory operation and data delivery, the computational units and the local buffers are grouped into 16 Processing Lanes (PL), and each PL is capable of handling one neuron's activation. A Near-Zero Approximation Unit (NZAU) is designed and integrated as a pre-processing unit, reducing the near-zero multiplications under a specified threshold. The threshold can be adjusted to match a specific model, as the data input and the parameters of the model may affect the tolerance of error. A optimal threshold can be determined by increasing it gradually while keeping the error rate of the model, and thus maximum NZ-Sparsity can be achieved. With the NZAU and local buffers, the implemented accelerator can perform near-zero sparse matrix-vector multiplications without the involvement of computational units in most of the time.

\subsection{Near-Zero Approximation Unit (NZAU)}
The Near-Zero Approximation Unit is implemented to support the aforementioned mechanism of approximation. To detect the leading zeros of the operands, 16-bit Leading-Zero-Counting (LZC) units are employed in the design. This LZC unit adopts a shared-carry propagate architecture and features small area and high energy efficiency \cite{LZC} , thus reducing the overhead of NZAU. 
	
The block diagram of the NZAU is given in Fig. \ref{fig:nzau}. Two 16-bit fixed point operands are first transferred to absolute values and then fed to Leading Zero Counting (LZC) units to calculate the leading zeros. After that, the leading zero counts are summed up and compared with a 5-bit threshold, which is adjustable to control the constraint of the near-zero approximation. If the total leading zeros exceed the threshold value, the corresponding multiplication is approximated to 0, and the operands are discarded. Otherwise, the approximation is not acceptable, and the operands will be passed to a PL for further processing. To support parallel processing of the 16 input and output neurons, 17 data words are fetched in every cycle, including 16 weights and 1 shared neuron input. In this case, only 17 LZC units are employed in this design to minimize the area and power cost of NZAU.  

	\subsection{Processing Lane}
		\begin{figure}[t]
			\centering
			
			\subfloat[Example of matrix-vector multiplication with near-zero sparsity]{%
				\includegraphics[width=0.4\textwidth]{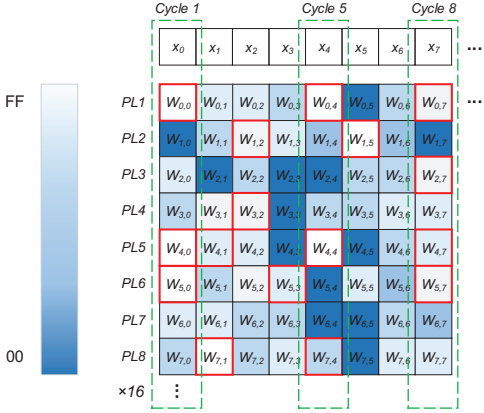}%
				\label{fig:compute_a}%
			}\qquad
			\subfloat[Data buffering for PL1 in the example]{%
				\includegraphics[width=0.4\textwidth]{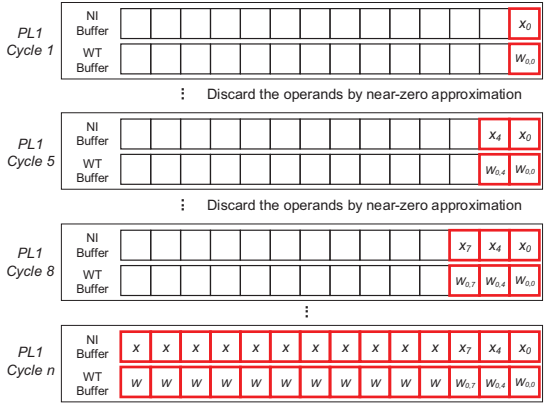}%
				\label{fig:compute_b}%
			}
			
			\caption{Data buffering for each Processing Lane, two data buffers for Neuron Inputs (NI) and Weights (WT) are included in each PL to store the pairs of operands }
		\end{figure}

The Processing Lane is a channel of processing one neuron's activation. The structure of the PL is shown in Fig. \ref{fig:pl}, consisting of two data buffers for the operands, two 256-bit operand registers, an array of 16 fixed-point multipliers and a adder tree. The operations within a PL can be divided into two phases: 1) data buffering; 2) computation.
	
Data buffering of the PL is triggered by the corresponding enabling signal ($data\_ld$ in Fig. \ref{fig:pl}) from NZAU. In every cycle, operands that can not be discarded by the NZAU are fed into the two operands buffers separately. The data buffer is designed to store the 16-bit input data sub-words and assemble them into a 256-bit full data word. When the $data\_ld$ signal is generated from NZAU, corresponding PL stores the pair of operands in the two buffers and updates the 4-bit counter. After the 256-bit full data word is assembled, the counter overflow flag enables the operand registers to save the data word, isolating the data buffering from computation.  
	
In Fig. \ref{fig:compute_a}, an example of matrix-vector multiplication with near-zero sparsity is given for illustration. In every cycle, one column of 16 weights and one neuron input are read from the two memories. The red boxes indicate the multiplications that need to be performed, while other multiplications are discarded due to the near-zero approximation. Operands that are stored in the data buffers of PL1 are shown in Fig. \ref{fig:compute_b}. The first 8 cycles contains only 3 multiplications that need processing, so only in cycle 1, 5 and 8, the corresponding pair of operands are loaded to the buffer. After $n$ cycles, the buffers are filled with 16 operands, then the 256-bit data word is flushed out and registered for computation. Thanks to the adjustable threshold, the near-zero sparsity can be enlarged to decrease the duty cycle of computational units. Therefore, the computational units can be clock-gated in most of the time, leading to significant reduction of the energy consumption.

\begin{table*}[]
	\centering
	\caption{Comparison with existing platforms}
	\label{tab_sum}
    \includegraphics[width=\textwidth]{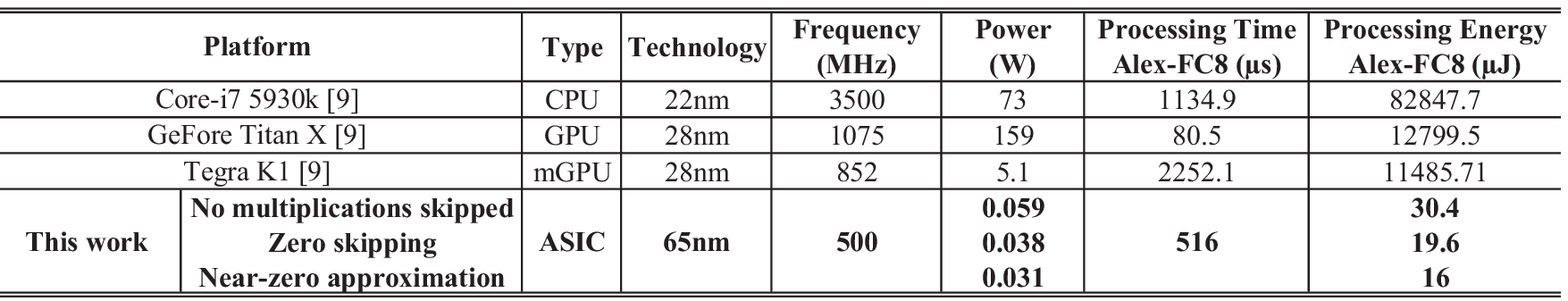}
\end{table*}

\section{Evaluations}

\begin{figure}[!t]
	\centering
	\includegraphics[width=0.45\textwidth]{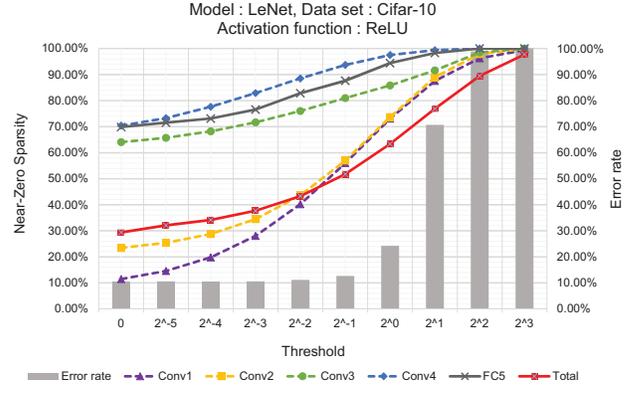}
	\caption{Test of near-zero approximation on LeNet-5: sparsity of the entire model is enlarged from 29.4\% to 63.35\% with negligible increase in error rate}
	\label{fig:lenet}
\end{figure}

\begin{figure}[!t]
	\centering
	\includegraphics[width=0.45\textwidth]{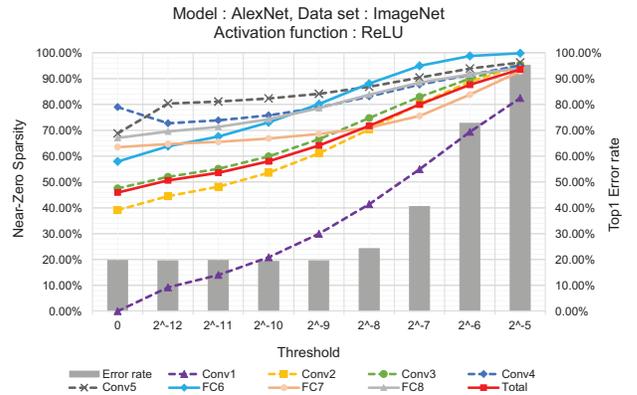}
	\caption{Test of near-zero approximation on Alexnet: sparsity of the entire model is enlarged from 45.9\% to 64.1\% with negligible increase in error rate}
	\label{fig:alexnet}
\end{figure}
\subsection{Validation of Near-Zero Approximation}


To validate the effectiveness of the proposed near-zero approximation, we apply the near-zero approximation to two classical Convolutional Neural Networks: LeNet-5 and Alexnet, as they are more typical to represent the real case. Various threshold values are employed to alter the NZ-Sparsity so as to test the models' error rate under different approximation levels.

The experimental results of LeNet-5 is illustrated in Fig. \ref{fig:lenet}, due to dense image data in Cifar-10, the first two layers have a relatively small sparsity lower than 25\% since they are close to the original data. However, as the layer goese deeper, the sparsity increases dramatically. It is mainly because that ReLU activation has an accumulating effect of producing zero-valued data. In other words, the zero-valued data produced by ReLU activation in the previous layer tend to result in more zero-valued data in current layer. Though the model has a zero-valued sparsity of 29.40\%, it can be further enlarged by applying the proposed near-zero approximation. When the model is constraint by the threshold of {2$^{-1}$}, the NZ-Sparsity of the whole model can reach 63.35\% with only 0.58\% increase of the error rate. Compared with zero-valued sparsity, the near-zero approximation reduces total number of multiplications by 1.92X. Similar results are found in testing the Alexnet, as shown in Fig. \ref{fig:alexnet}. The sparsity of Alexnet can be enlarged from 45.9\% to 64.1\% with negligible loss of accuracy, corresponding to 1.51X reduction of the total multiplications. 
	
\subsection{Evaluation of the accelerator}
The proposed accelerator was synthesized in UMC 65nm Low Leakage Process by Synopsys Design Compiler. Switching activity of processing the Alexnet-FC8 layer was annotated in Synopsys Prime Power to estimate the power consumption of the accelerator. Operating at 500 MHz, the entire accelerator consumes 59 mW without optimization of sparse data at 500 MHz, and the number drops to 38 mW when the zero-skipping is used. When the near-zero approximation is applied without sacrificing the accuracy, the power consumption is reduced to 31 mW, corresponding to 18.4\% further improvement. Table \ref{tab_sum} summaries this work and compares with other existing mainstream platforms \cite{EIE}. It can be seen that the proposed accelerator is $>$ 2X and $>$ 4X faster than the high performance CPU for PC and the mobile GPU when processing the last fully-connected layer of Alexnet. Considering the total energy of executing the task, our accelerator exhibits 717X energy reduction than mobile GPU Tegra K1. 


	
\section{Conclusion}
	
There exists a large sparsity in the Deep Neural Network model, which can be utilized to optimize the DNN for IoT devices. But the sparsity is only restricted to zero-valued data. In this work, we extend the zero-valued sparsity to near-zero valued sparsity (NZ-Sparisty) and proposed a near-zero approximation technique to reduce the near-zero multiplications in learning models. Dedicated hardware of the approximation scheme, named Near-Zero Approximation Unit, is designed and integrated in the accelerator for energy-efficient processing of DNNs. Evaluated in the UMC 65nm process, the 500 MHz accelerator is $>$ 4X faster than the mobile GPU Tegra K1 in processing the fully-connected layer FC8 of Alexnet, while consuming 717X less energy.



\bibliographystyle{IEEEtran}
\bibliography{IEEEabrv,./bib}

\begin{thebibliography}{10}
\providecommand{\url}[1]{#1}
\csname url@samestyle\endcsname
\providecommand{\newblock}{\relax}
\providecommand{\bibinfo}[2]{#2}
\providecommand{\BIBentrySTDinterwordspacing}{\spaceskip=0pt\relax}
\providecommand{\BIBentryALTinterwordstretchfactor}{4}
\providecommand{\BIBentryALTinterwordspacing}{\spaceskip=\fontdimen2\font plus
\BIBentryALTinterwordstretchfactor\fontdimen3\font minus
  \fontdimen4\font\relax}
\providecommand{\BIBforeignlanguage}[2]{{%
\expandafter\ifx\csname l@#1\endcsname\relax
\typeout{** WARNING: IEEEtran.bst: No hyphenation pattern has been}%
\typeout{** loaded for the language `#1'. Using the pattern for}%
\typeout{** the default language instead.}%
\else
\language=\csname l@#1\endcsname
\fi
#2}}
\providecommand{\BIBdecl}{\relax}
\BIBdecl

\bibitem{ISCAS14}
Z.~Chen and \emph{et al.}, ``A fast deep learning system using gpu,'' in
  \emph{2014 IEEE International Symposium on Circuits and Systems (ISCAS)},
  June 2014, pp. 1552--1555.

\bibitem{icml2013}
A.~Coates and \emph{et al.}, ``Deep learning with cots hpc systems,'' in
  \emph{Proceedings of the 30th International Conference on Machine Learning
  (ICML-13)}, May 2013, pp. 1337--1345.

\bibitem{ISCA10}
Chakradhar and \emph{et al.}, ``A dynamically configurable coprocessor for
  convolutional neural networks,'' in \emph{Proceedings of the 37th Annual
  International Symposium on Computer Architecture}, ser. ISCA '10, New York,
  NY, USA, 2010, pp. 247--257.

\bibitem{CVPR14}
V.~Gokhale and \emph{et al.}, ``A 240 g-ops/s mobile coprocessor for deep
  neural networks,'' in \emph{2014 IEEE Conference on Computer Vision and
  Pattern Recognition Workshops}, June 2014, pp. 696--701.

\bibitem{ASPDAC16}
S.~Venkataramani and \emph{et al.}, ``Efficient embedded learning for iot
  devices,'' in \emph{2016 21st Asia and South Pacific Design Automation
  Conference (ASP-DAC)}, Jan 2016, pp. 308--311.

\bibitem{alexnet}
A.~Krizhevsky and \emph{et al.}, ``Imagenet classification with deep
  convolutional neural networks,'' in \emph{Advances in neural information
  processing systems}, 2012, pp. 1097--1105.

\bibitem{hanNIPS15}
S.~Han and \emph{et al.}, ``Learning both weights and connections for efficient
  neural network,'' in \emph{Advances in Neural Information Processing
  Systems}, 2015, pp. 1135--1143.

\bibitem{eyeriss}
Y.~H. Chen and \emph{et al.}, ``Eyeriss: An energy-efficient reconfigurable
  accelerator for deep convolutional neural networks,'' in \emph{2016 IEEE
  International Solid-State Circuits Conference (ISSCC)}, Jan 2016, pp.
  262--263.

\bibitem{EIE}
S.~Han and \emph{et al.}, ``Eie: Efficient inference engine on compressed deep
  neural network,'' in \emph{2016 ACM/IEEE 43rd Annual International Symposium
  on Computer Architecture (ISCA)}, June 2016, pp. 243--254.

\bibitem{VLSI16}
B.~Moons and M.~Verhelst, ``A 0.3-2.6 tops/w precision-scalable processor for
  real-time large-scale convnets,'' in \emph{2016 IEEE Symposium on VLSI
  Circuits (VLSI-Circuits)}, June 2016, pp. 1--2.

\bibitem{lenet}
Y.~LeCun and \emph{et al.}, ``Gradient-based learning applied to document
  recognition,'' \emph{Proceedings of the IEEE}, vol.~86, no.~11, pp.
  2278--2324, 1998.

\bibitem{HYX_SOCC16}
Y.~Huan and \emph{et al.}, ``A multiplication reduction technique with
  near-zero approximation for embedded learning in iot devices,'' in \emph{2016
  29th IEEE International System-on-Chip Conference (SOCC)}, Sept 2016.

\bibitem{LZC}
G.~Dimitrakopoulos and \emph{et al.}, ``Low-power leading-zero counting and
  anticipation logic for high-speed floating point units,'' \emph{IEEE
  Transactions on Very Large Scale Integration (VLSI) Systems}, vol.~16, no.~7,
  pp. 837--850, July 2008.

\end{thebibliography}

\end{document}